\documentstyle[12pt]{article}
\makeatletter
\def\thesection{\arabic{section}}
\@addtoreset{equation}{section}
\def\theequation{\thesection.\arabic{equation}}
\def\appendix{\par
\setcounter{section}{0}
\setcounter{subsection}{0}
\def\thesection{\Alph{section}}}
\makeatother
\makeatletter
\def\abstract#1{\long\def\@abstract{#1}}%
\def\@abstract{}%
\let\@oldmaketitle=\@maketitle%
\def\@maketitle{%
\@oldmaketitle%
\begin{center}\large\bf Abstract\end{center}%
\begin{quotation}\@abstract\end{quotation}%
\vskip 1.5em}%
\makeatother
\makeatletter
\def\eqnarray{%
\stepcounter{equation}%
\let\@currentlabel=\theequation
\global\@eqnswtrue
\global\@eqcnt\z@
\tabskip\@centering
\let\\=\@eqncr
$$\halign to \displaywidth\bgroup\@eqnsel\hskip\@centering
$\displaystyle\tabskip\z@{##}$&\global\@eqcnt\@ne
\hfil$\displaystyle{{}##{}}$\hfil
&\global\@eqcnt\tw@$\displaystyle\tabskip\z@{##}$\hfil
\tabskip\@centering&\llap{##}\tabskip\z@\cr}
\makeatother
\newcommand{\fxi}{\mbox{\boldmath$\xi$}}

\newcommand{\ffp}{\mbox{\boldmath$\varphi$}}

\newcommand{\ffz}{\mbox{\boldmath$z$}}

\newcommand{\delt}{\Delta t}
\newcommand{\tr}{{\rm tr}}

\newcommand{\bra}[1]{{\langle{#1}\vert}}
\newcommand{\ket}[1]{{\vert{#1}\rangle}}
\newcommand{\braket}[2]{{\langle{#1}\vert{#2}\rangle}}
\newcommand{\kansu}[2]{{{#1}\!\left({#2}\right)}}
\newcommand{\kakko}[1]{{\left({#1}\right)}}
\newcommand{\vp}{\varphi}
\newcommand{\ve}{\varepsilon}
\newcommand{\wa}[2]{\sum^{#1}_{#2}}
\begin{document}

\title{\sl Multi-Periodic Coherent States\\and
\\ the WKB-Exactness {\rm II}\\
{\large\sl ``Non-compact Case and
Classical theories Revisited''}
}
\author{
  Kazuyuki FUJII
  \thanks{e-mail address : fujii@yokohama-cu.ac.jp}\\
  Department of Mathematics, Yokohama City University,\\
  Yokohama 236, Japan\\\\
  Kunio FUNAHASHI
  \thanks{e-mail address : fnhs1scp@mbox.nc.kyushu-u.ac.jp}\\
  Department of Physics, Kyushu University,\\
  Fukuoka 812-81, Japan}
\date{October, 1996}

\abstract{
We show that the WKB approximation gives the exact result
in the trace formula of ``$CQ^N$'', which is the non-compact counterpart
of $CP^N$, in terms of the ``multi-periodic'' coherent state.
We revisit the symplectic 2-forms on $CP^N$ and $CQ^N$ and,
especially, construct that on $CQ^N$ with the unitary form.
We also revisit the exact calculation of the classical patition
functions of them.
}
\maketitle\thispagestyle{empty}
\newpage

\section{Introduction}\label{sec:jo}

Recently there have been many discussions~\cite{Stone,Rajeev,PCS,Blau}
on the system in which the WKB approximation gives the exact result
(We call this fact as the ``WKB-exactness'') in path integral
in connection with the Duistermaat-Heckman (DH) theorem~\cite{DH,Atiyah}.
As for our work we have shown that $SU(2)$ spin and its non-compact
counterpart $SU(1,1)$ are the WKB-exact in the path integral
formulas constructed in terms of the spin coherent state with discrete
time method~\cite{FKSF1}.
We have also shown~\cite{WESHS,FKS} the WKB-exactness of their extensions;
some unitary representations of $U(N+1)$ and $U(N,1)$,
whose phase spaces are $CP^N$ and $CQ^N$ (the definition
is given in section \ref{sec:koten}) respectively in terms of
the generalized coherent states~\cite{Perelomov} by the help of
the Schwinger boson method~\cite{Schwinger}.

Then there arises a question whether the WKB-exactness holds in some
special coherent states.
It is important to investigate the WKB-exactness of the same
systems in terms of other coherent states.
With respect to the $SU(2)$ spin, there is another expression,
the Nielsen-Rohrlich formula~\cite{NR},
which is constructed in terms of the periodic coherent state~\cite{TK}.
We have shown the WKB-exactness of the Nielsen-Rohrlich formula
{}~\cite{FKNS} and its extension to $U(N+1)$ in terms of
the ``multi-periodic'' coherent state~\cite{KFKF}
although their handling is more delicate than that of the generalized
coherent state cases.
In the fisrt half of this paper, we discuss the WKB-exactness
in the trace formula of some representation of $U(N,1)$.

In the latter half of this paper we revisit classical systems
whose phase spaces are $CP^N$ and $CQ^N$ respectivly.
We reconsider the symplectic 2-forms on $CP^N$  and $CQ^N$.
It is matter of course that they are easily constructed as we perform.
As an alternative way, embedding $CQ^N$ to $CP(l^2(C))$ skillfully,
we construct the symplectic 2-form on $CQ^N$ by the pullback of
$CP(l^2(C))$.
This embedding gives the unitary expression of $CQ^N$.
Thus we can comprehend the symplectic 2-forms on $CP^N$ and $CQ^N$
in a unified manner.
Next we revisit the exact calculations of the classical
partition function of $CP^N$ and $CQ^N$.
The applications of DH theorem on classical systems had
been already made to the flag manifold~\cite{RFP}.
However the exact calculation which should be compared with
the result of the stationary phase approximation is very
complicated.
We make an attempt to calculate the partition functions of
$CP^N$ and $CQ^N$ by the direct calculation and
by lifting to the Gaussian integral forms.
The comparison between these two methods gives us  deep
understanding of the DH theorem.
We also consider the construction of the partition function
of $CQ^N$ by the pullback of $CP(l^2(C))$.

\section{Trace Formula and the WKB-Exactness of Path Integral}
\label{sec:multi:noncompact}

In this section we show the WKB-exactness of a representation of $CQ^N$.
In subsection \ref{sec:multi:kousei} we construct the trace
formula by the help of the Schwinger boson method~\cite{Schwinger}.
We calculate the trace formula exactly in subsection
\ref{sec:multi:genmitsu}.
Then in subsection \ref{sec:multi:wkb} we make the WKB
approximation and compare the result with that of the
exact calculation.

\subsection{Construction of the Trace Formula}
\label{sec:multi:kousei}

$u(N,1)$ algebra is defined by
\begin{eqnarray}
  \label{ncp:daisuteigi}
  &&
  [E_{\alpha\beta},E_{\gamma\delta}]
  =
  \eta_{\beta\gamma}E_{\alpha\delta}
  -\eta_{\delta\alpha}E_{\gamma\beta}\ ,
  \nonumber\\
  &&
  \eta_{\alpha\beta}
  =
  \kansu{\rm diag}{1,\cdots,1,-1}\ ,\
  \kakko{\alpha,\beta,\gamma,\delta = 1,\cdots,N+1}\ ,
\end{eqnarray}
with a subsidiary condition
\begin{equation}
  -\wa{N}{\alpha=1}E_{\alpha\alpha}+E_{N+1,N+1}=K\ ,\
  \kakko{K=N,N+1,\cdots}\ .
\end{equation}
We identify these generators with creation and annihilation operators
of harmonic oscillator:
\begin{equation}
  \matrix{
    &E_{\alpha\beta}
    =a_\alpha^\dagger a_\beta,&
    E_{\alpha,N+1}
    =a_\alpha^\dagger a_{N+1}^\dagger
    \cr
    &E_{N+1,\alpha}
    =a_{N+1}a_\alpha,&
    E_{N+1,N+1}
    =a_{N+1}^\dagger a_{N+1}+1
    \cr
    }\ ,
\end{equation}
where $a$, $a^\dagger$ satisfy
\begin{eqnarray}
  [a_\alpha,a_\beta^\dagger]=1\ ,\
  [a_\alpha,a_\beta]=[a_\alpha^\dagger,a_\beta^\dagger]=0\ ,
  \nonumber\\
  \kakko{\alpha,\beta=1,2,\cdots,N+1}\ ,
\end{eqnarray}
and the Fock space is
\begin{eqnarray}
  \left\{
    \ket{n_1,\cdots,n_{N+1}}\vert n_1,n_2,\cdots,n_{N+1}=0,1,2,\cdots
  \right\}\ ,
  \nonumber\\
  \ket{n_1,\cdots,n_{N+1}}
  \equiv
  {1\over\sqrt{n_1!\cdots n_{N+1}!}}
  \kakko{a_1^\dagger}^{n_1}\cdots\kakko{a_{N+1}^\dagger}^{n_{N+1}}
  \ket{0,0,\cdots,0}\ .
\end{eqnarray}
The representation space is
\begin{equation}
  1_K
  \equiv
  \wa{\infty}{\left\{ n\right\}=0}
  \ket{n_1,\cdots,n_N,K-1+\sum^N_{\alpha=1}n_\alpha}
  \bra{n_1,\cdots,n_N,K-1+\sum^N_{\alpha=1}n_\alpha}\ ,
\end{equation}
where
\begin{equation}
  \sum^\infty_{\left\{ n\right\}=0}
  \equiv
  \wa{\infty}{n_1=0}
  \wa{\infty}{n_2=0}
  \cdots
  \wa{\infty}{n_N=0}\ .
\end{equation}

The ``multi-periodic'' coherent state is defined by
\begin{eqnarray}
  \label{ncp:csteigi}
  \ket{\ffp}
  &\equiv&
  \ket{\vp_1,\cdots,\vp_N}
  \nonumber\\
  &\equiv&
    {1\over\kakko{2\pi}^{N/2}}
  \wa{\infty}{\left\{ m\right\}=0}
  \sqrt{\kakko{K-1}!\over m_1!\cdots m_N!
    \kakko{K-1+\wa{N}{\alpha=1}m_\alpha}!}
  \nonumber\\
  &&\times
  \prod^N_{\alpha=1}
  \left\{
    e^{-im_\alpha\vp_\alpha}
    \kakko{E_{\alpha,N+1}}^{m_\alpha}
  \right\}
  \ket{0,\cdots,0,K-1}
  \nonumber\\
  &=&
  {1\over\kakko{2\pi}^{N/2}}
  \wa{\infty}{\left\{ m\right\}=0}
  e^{-i\wa{N}{\alpha=1}m_\alpha\vp_\alpha}
  \ket{m_1,\cdots,m_N,K-1+\wa{N}{\alpha=1}m_\alpha}\ ,
\end{eqnarray}
which has the periodic property
\begin{equation}
  \ket{\vp_1,\cdots,\vp_\alpha+2n\pi,\cdots,\vp_N}
  =
  \ket{\vp_1,\cdots,\vp_\alpha,\cdots,\vp_N}\ ,
\end{equation}
for each $\vp_\alpha(\alpha=1,\cdots,N)$.
These states satisfy the resolution of unity
\begin{equation}
  \label{ncp:kanzensei}
  \int^{2\pi}_0
  \prod^N_{\alpha=1}d\vp_\alpha
  \ket{\ffp}\bra{\ffp}
  =
  1_K\ .
\end{equation}
Their inner product is
\begin{equation}
  \label{ncp:naiseki}
  \braket{\ffp}{\ffp^\prime}
  =
  {1\over\kakko{2\pi}^N}
  \wa{\infty}{\left\{ m\right\}=0}
  e^{i\wa{N}{\alpha=1}m_\alpha
    \kakko{\vp_\alpha-\vp_\alpha^\prime}}\ .
\end{equation}

We adopt the Hamiltonian
\begin{eqnarray}
  \label{ncq:hamiltonian}
  \hat H
  &\equiv&
  \wa{N+1}{\alpha=1}c_\alpha E_{\alpha\alpha}
  \nonumber\\
  &=&
  \wa{N}{\alpha=1}\mu_\alpha E_{\alpha\alpha}+Kc_{N+1}\ ,
\end{eqnarray}
where
\begin{equation}
  \mu_\alpha \equiv c_\alpha+c_{N+1}\ .
\end{equation}
Its matrix element is
\begin{equation}
  \bra{\ffp}\hat H\ket{\ffp^\prime}
  =
  \kakko{
    \wa{N}{\alpha=1}
      \mu_\alpha{\partial\over\partial(i\vp_\alpha)}
    +Kc_{N+1}}
  \braket{\ffp}{\ffp^\prime}\ .
\end{equation}

To construct a path integral formula, we rewrite the inner product
(\ref{ncp:naiseki}), with the help of the formula
\begin{eqnarray}
  \label{ncp:saishonokoushiki}
  \wa{m_1}{m=m_0}e^{im\vp}\kansu{f}{m}
  &=&
  \wa{\infty}{n=-\infty}
  \int^{m_1+\ve_0}_{m_0-\ve_0}dp
  e^{ip\kakko{\vp+2n\pi}}
  \kansu{f}{p}\ ,
  \nonumber\\
  \kakko{0<\ve_0<1}\ ,
  \nonumber\\
  &&{\rm for}\ m_0,m_1\in{\bf Z}\ ,
\end{eqnarray}
to
\begin{eqnarray}
  \braket{\ffp}{\ffp^\prime}
  &=&
  \prod^N_{\alpha=1}
  \left[
    {1\over2\pi}\wa{\infty}{n_\alpha=-\infty}
    \int^\infty_{-\ve_0}dp_\alpha
    e^{ip_\alpha\kakko{\vp_\alpha-\vp_\alpha^\prime+2n_\alpha\pi}}
  \right]
  \nonumber\\
  &=&
  \wa{\infty}{\left\{ n_\alpha\right\}=-\infty}
  \int^\infty_{-\ve_0}
  \prod^N_{\alpha=1}{dp_\alpha\over2\pi}
  e^{i\wa{N}{\alpha=1}p_\alpha
    \kakko{\vp_\alpha-\vp_\alpha^\prime+2n_\alpha\pi}}\ .
\end{eqnarray}
We make a change of variables such that
\begin{equation}
  \tilde p_\alpha = p_\alpha+{1\over2}\ ,\
  \ve = -\ve_0+{1\over2}\ ,\ \kakko{0\le\ve<{1\over2}}\ ,
\end{equation}
to obtain
\begin{equation}
  \label{ncp:naiseki:gutai}
  \braket{\ffp}{\ffp^\prime}
  =
  \wa{\infty}{\left\{ n_\alpha\right\}=-\infty}
  \int^\infty_\ve
  \prod^N_{\alpha=1}d\tilde p_\alpha
  e^{i\wa{N}{\alpha=1}
    \kakko{\tilde p_\alpha-1/2}
    \kakko{\vp_\alpha-\vp_\alpha^\prime+2n_\alpha\pi}}\ .
\end{equation}

The Feynman kernel is defined by
\begin{eqnarray}
  \label{ncp:kernel:teigi}
  \kansu{K}{\ffp_F,\ffp_I;T}
  &\equiv&
  \bra{\ffp_F}e^{-i\hat HT}\ket{\ffp_I}
  \nonumber\\
  &=&
  \lim_{M\to\infty}
   \bra{\ffp_F}\kakko{1-i\delt\hat H}^M\ket{\ffp_I}\ ,\
   \kakko{\delt\equiv T/M}\ ,
\end{eqnarray}
where $\ffp_I(\ffp_F)$ is the initial (final) state and $T$
is time interval.
By inserting the resolution of unity (\ref{ncp:kanzensei})
between each $(1-i\delt\hat H)$ of its product, the kernel becomes
\begin{eqnarray}
  \label{ncp:kernel:chuuto}
  \kansu{K}{\ffp_F,\ffp_I;T}
  &=&
  \lim_{M\to\infty}
  \int^{2\pi}_0
  \prod^{M-1}_{i=1}\prod^N_{\alpha=1}
  {\kansu{d\vp_\alpha}{i}\over2\pi}
  \nonumber\\
  &&\times
  \prod^M_{j=1}
  \bra{\kansu{\ffp}{j}}
  \kakko{1-i\delt\hat H}
  \ket{\kansu{\ffp}{j-1}}
  \Bigg\vert^{\kansu{\ffp}{M}=\ffp_F}_{\kansu{\ffp}{0}=\ffp_I}\ .
\end{eqnarray}
Each term of the product becomes
\begin{eqnarray}
  &&\bra{\kansu{\ffp}{j}}
  \kakko{1-i\delt\hat H}
  \ket{\kansu{\ffp}{j-1}}
  \nonumber\\
  &=&
  \wa{\infty}{\left\{\kansu{n}{j}\right\}=-\infty}
  \int^\infty_\ve
  \prod^N_{\alpha=1}d\kansu{p_\alpha}{j}
  e^{
    i\wa{N}{\alpha=1}
    \kakko{\kansu{p_\alpha}{j}+1/2}
    \kakko{\kansu{\vp_\alpha}{j}-\kansu{\vp_\alpha}{j-1}
      +2\kansu{n_\alpha}{j}\pi}
    }
  \nonumber\\
  &&\times
  \Bigg[
  1-i\delt\bigg\{\wa{N}{\alpha=1}\mu_\alpha
  \kakko{\kansu{p_\alpha}{j}-{1\over2}}+Kc_{N+1}\bigg\}
  \Bigg]
  \nonumber\\
  &=&
  \wa{\infty}{\left\{\kansu{n}{j}\right\}=-\infty}
  \int^\infty_\ve
  \prod^N_{\alpha=1}d\kansu{p_\alpha}{j}
  e^{
    i\wa{N}{\alpha=1}
    \kakko{\kansu{p_\alpha}{j}+1/2}
    \kakko{\kansu{\vp_\alpha}{j}-\kansu{\vp_\alpha}{j-1}
      +2\kansu{n_\alpha}{j}\pi}
    }
  \nonumber\\
  &&\times
  \exp\Bigg[
  -i\delt\bigg\{
    \wa{N}{\alpha=1}\mu_\alpha
    \kakko{\kansu{p_\alpha}{j}-{1\over2}}
    +Kc_{N+1}\bigg\}
  \Bigg]\ ,
\end{eqnarray}
where the primes of $p$'s have been omitted and
the explicit form of the inner product (\ref{ncp:naiseki:gutai})
has been put in the first equality and $\kansu{O}{\delt^2}$ terms,
which vanish in the $M\to\infty$ limit, have been omitted
in the second equality.
Thus the kernel (\ref{ncp:kernel:chuuto}) is
\begin{eqnarray}
  \kansu{K}{\ffp_F,\ffp_I;T}
  &=&
  \lim_{M\to\infty}
  e^{-iKc_{N+1}T}
  \prod^N_{\alpha=1}
  \Bigg[
  \kakko{\prod^M_{i=1}
    \wa{\infty}{\kansu{n_\alpha}{i}=-\infty}}
  \int^{2\pi}_0
  \prod^{M-1}_{i=1}
  {d\kansu{\vp_\alpha}{i}\over2\pi}
  \int^\infty_\ve
  \prod^M_{j=1}
  d\kansu{p_\alpha}{j}
  \nonumber\\
  &&\times
  \exp\bigg[
  i\wa{M}{k=1}
  \kakko{\kansu{p_\alpha}{k}-{1\over2}}
  \kakko{\kansu{\vp_\alpha}{k}-\kansu{\vp_\alpha}{k-1}
    +2\kansu{n_\alpha}{k}\pi-\delt\mu_\alpha}
  \bigg]
  \Bigg]
  \nonumber\\
  &=&
  \lim_{M\to\infty}
  e^{-iKc_{N+1}T}
  \prod^N_{\alpha=1}
  \Bigg[
  \wa{\infty}{\kansu{n_\alpha^\prime}{M}=-\infty}
  \int^\infty_{-\infty}
  \prod^{M-1}_{i=1}
  {d\kansu{\vp_\alpha^\prime}{i}\over2\pi}
  \int^\infty_0
  \prod^M_{i=1}
  d\kansu{p_\alpha}{j}
  \nonumber\\
  &&\times
  \exp
  \bigg[
  i\wa{M}{k=1}
  \kakko{\kansu{p_\alpha}{k}-{1\over2}}
  \kakko{\kansu{\vp_\alpha^\prime}{k}
    -\kansu{\vp_\alpha^\prime}{k-1}
    -\delt\mu_\alpha}
  \bigg]
  \Bigg]
  \Bigg\vert^{
    \kansu{\vp_\alpha^\prime}{M}
    =\kakko{\vp_F}_\alpha
    +2\kansu{n_\alpha^\prime}{M}\pi
    }_{
    \kansu{\vp_\alpha^\prime}{0}
    =\kakko{\vp_I}_\alpha
    }\ ,
\end{eqnarray}
where changing variables
\begin{eqnarray}
  &&
  \kansu{n_\alpha^\prime}{k}
  =
  \wa{k}{l=1}\kansu{n_\alpha}{l}\ ,
  \nonumber\\
  &&
  \kansu{\vp_\alpha^\prime}{k}
  =
  \kansu{\vp_\alpha}{k}
  +2\kansu{n^\prime_\alpha}{k}\pi\ ,
\end{eqnarray}
which leads to
\begin{eqnarray}
  &&\kansu{\vp_\alpha}{k}
  -\kansu{\vp_\alpha}{k-1}
  +2\kansu{n_\alpha}{k}\pi
  =
  \kansu{\vp_\alpha^\prime}{k}
  -\kansu{\vp_\alpha^\prime}{k-1}\ ,
  \nonumber\\
  &&
  \wa{\infty}{\kansu{n_\alpha}{i}=-\infty}
  \int^{
    2\kakko{\kansu{n_\alpha}{i}+1}\pi
    }_{
    2\kansu{n_\alpha}{i}\pi
    }
  d\kansu{\vp_\alpha}{i}
  =
  \int^\infty_{-\infty}
  d\kansu{\vp_\alpha^\prime}{i}\ ,
\end{eqnarray}
has been made.
To compare the kernel with that of the compact case,
we further put
\begin{equation}
  \kansu{p_\alpha}{k}
  =
  \kansu{\cosh}{\kansu{\theta_\alpha}{k}}-1\ .
\end{equation}
Finally we obtain
\begin{eqnarray}
  \label{ncp:kernel:gutai}
  \kansu{K}{\ffp_F,\ffp_I;T}
  &=&
  \lim_{M\to\infty}
  e^{-iKc_{N+1}T}
  \prod^N_{\alpha=1}
  \Bigg[
  e^{i{3\over2}\mu_\alpha T}
  \wa{\infty}{\kansu{n_\alpha}{M}=-\infty}
  e^{-i{3\over2}\kakko{\kansu{\vp_\alpha}{M}-\kansu{\vp_\alpha}{0}}}
  \nonumber\\
  &&\times
  \int^\infty_{-\infty}
  \prod^{M-1}_{i=1}
  {d\kansu{\vp_\alpha}{i}\over2\pi}
  \int^\infty_0
  \prod^M_{j=1}
  \kansu{\sinh}{\kansu{\theta_\alpha}{j}}
  d\kansu{\theta_\alpha}{j}
  \nonumber\\
  &&\times
  \exp
  \bigg[
  i\wa{M}{k=1}
  \kansu{\cosh}{\kansu{\theta_\alpha}{k}}
  \kakko{\kansu{\vp_\alpha}{k}
    -\kansu{\vp_\alpha}{k-1}
    -\delt\mu_\alpha}
  \bigg]
  \Bigg]
   \Bigg\vert^{
    \kansu{\vp_\alpha}{M}
    =\kakko{\vp_F}_\alpha
    +2\kansu{n_\alpha}{M}\pi
    }_{
    \kansu{\vp_\alpha}{0}
    =\kakko{\vp_I}_\alpha
    }\ ,
\end{eqnarray}
in the $\theta$-expression.

The trace formula is defined by
\begin{eqnarray}
  Z
  &\equiv&
  \int^{2\pi}_0
  \prod^N_{\alpha=1}d\vp_\alpha
  \bra{\ffp}e^{-i\hat HT}\ket{\ffp}
  \nonumber\\
  &\equiv&
  \int^{2\pi}_0
  \prod^N_{\alpha=1}
  d\vp_\alpha
  \kansu{K}{\ffp_F,\ffp_I;T}\ .
\end{eqnarray}
{}From (\ref{ncp:kernel:gutai}), the explicit form is
\begin{eqnarray}
  \label{ncp:ato}
  Z
  &=&
  \lim_{M\to\infty}
  e^{-iKc_{N+1}T}
  \prod^N_{\alpha=1}
  \Bigg[
  e^{i{3\over2}\mu_\alpha T}
  \wa{\infty}{n_\alpha=-\infty}
  e^{-i3n_\alpha\pi}
  \nonumber\\
  &&\times
  \int^{2\pi}_0
  {d\kansu{\vp_\alpha}{M}\over2\pi}
  \int^\infty_{-\infty}
  \prod^{M-1}_{i=1}
  {d\kansu{\vp_\alpha}{i}\over2\pi}
  \int^\infty_0
  \prod^M_{j=1}
  \kansu{\sinh}{\kansu{\theta_\alpha}{j}}
  d\kansu{\theta_\alpha}{j}
  \nonumber\\
  &&\times
  \exp
  \bigg[
  i\wa{M}{k=1}
  \kansu{\cosh}{\kansu{\theta_\alpha}{k}}
  \kakko{
    \kansu{\vp_\alpha}{k}
    -\kansu{\vp_\alpha}{k-1}
    -\delt\mu_\alpha}
  \bigg]
  \Bigg]
  \Bigg\vert_{
    \kansu{\vp_\alpha}{M}
    =\kansu{\vp_\alpha}{0}+2n_\alpha\pi}\ .
\end{eqnarray}
This is the trace formula of $CQ^N$.

Comparing (\ref{ncp:ato}) with the trace formula of
$CP^N$~\cite{KFKF},
\begin{eqnarray}
  \label{ncp:cpnato}
  Z
  &=&
  e^{-iQc_{N+1}T
    +i{1\over N+1}\wa{N}{\alpha=1}2^{-\alpha+1}\mu_\alpha T}
  \wa{\infty}{n_1=-\infty}\cdots\wa{\infty}{n_N=-\infty}
  e^{-i{1\over N+1}\wa{N}{\alpha=1}2^{-\alpha+1}2n_\alpha\pi}
  \nonumber\\
  &&\times
  \lim_{M\to\infty}\prod^N_{\alpha=1}
  \left\{
    \int^{2\kakko{n_\alpha+1}\pi}_{2n_\alpha\pi}
    d\kansu{\vp_\alpha}{M}
  \right\}
  \int^\infty_{-\infty}
  \prod^{M-1}_{i=1}
  \prod^N_{\alpha=1}
  d\kansu{\vp_\alpha}{i}
  \nonumber\\
  &&\times
  \prod^M_{j=1}
  \Bigg\{
    \lambda^N
    \int^\pi_0
    \kakko{\sin^2{\kansu{\theta_1}{j}\over2}}^{N-1}
    \sin\kansu{\theta_1}{j}{d\kansu{\theta_1}{j}\over2\pi}
    \nonumber\\
    &&\times\cdots
    \int^\pi_0
    \sin^2{\kansu{\theta_{N-1}}{j}\over2}
    \sin\kansu{\theta_{N-1}}{j}
    {d\kansu{\theta_{N-1}}{j}\over2\pi}
    \int^\pi_0
    \sin\kansu{\theta_N}{j}
    {d\kansu{\theta_N}{j}\over2\pi}
  \Bigg\}
  \nonumber\\
  &&\times
  \exp
  \Bigg[
  2i\lambda\wa{M}{k=1}\wa{N}{\alpha=1}
  \kakko{\prod^{\alpha-1}_{\beta=1}
    \sin^2{\kansu{\theta_\beta}{k}\over2}}
  \cos^2{\kansu{\theta_\alpha}{k}\over2}
  \nonumber\\
  &&\times
  \kakko{\kansu{\vp_\alpha}{k}-\kansu{\vp_\alpha}{k-1}
    -\delt\mu_\alpha}
  \Bigg]
  \Bigg\vert_{\kansu{\vp_\alpha}{0}=\kansu{\vp_\alpha}{M}
    +2n_\alpha\pi}\ ,
  \nonumber\\
  &&
  \mu_\alpha\equiv c_\alpha-c_{N+1}\ ,\
  \lambda\equiv{Q\over2}+{1\over N+1}\ ,
\end{eqnarray}
we find that (\ref{ncp:ato}) consists of the product
of independent terms on $\alpha$ while (\ref{ncp:cpnato})
does not.

\subsection{The Exact Calculation}
\label{sec:multi:genmitsu}

We calculate the trace formula exactly.
In the $p$-expression the trace formula is
\begin{eqnarray}\label{gen:phyoji}
  Z
  &=&
  \lim_{M\to\infty}
  e^{-iKc_{N+1}T}
  \prod^N_{\alpha=1}
  \Bigg[
  \wa{\infty}{n_\alpha=-\infty}
  \int^{2\pi}_0{d\kansu{\vp_\alpha}{M}\over2\pi}
  \int^\infty_{-\infty}\prod^{M-1}_{i=1}
  {d\kansu{\vp_\alpha}{i}\over2\pi}
  \int^\infty_0\prod^M_{j=1}d\kansu{p_\alpha}{j}
  \nonumber\\
  &&\times
  \exp\bigg[i\wa{M}{k=1}
  \kakko{\kansu{p_\alpha}{k}-{1\over2}}
  \kakko{\kansu{\vp_\alpha}{k}-\kansu{\vp_\alpha}{k-1}
  -\delt\mu_\alpha}
  \bigg]
  \Bigg]\Bigg\vert_{\kansu{\vp_\alpha}{M}=\kansu{\vp_\alpha}{0}}
  \nonumber\\
  &=&
  \lim_{M\to\infty}
  e^{-iKc_{N+1}T}
  \prod^N_{\alpha=1}
  \Bigg[
  e^{{i\over2}\mu_\alpha T}
  \wa{\infty}{n_\alpha=-\infty}
  e^{-in_\alpha\pi}
  \nonumber\\
  &&\times
  \int^{2\pi}_0{d\kansu{\vp_\alpha}{M}\over2\pi}
  \int^\infty_{-\infty}
  \prod^{M-1}_{i=1}
  {d\kansu{\vp_\alpha}{i}\over2\pi}
  \int^\infty_0\prod^M_{j=1}d\kansu{p_\alpha}{j}
  \nonumber\\
  &&\times
  \exp\bigg[
  -i\wa{M-1}{k=1}
  \kakko{\kansu{p_\alpha}{k+1}-\kansu{p_\alpha}{k}}
  \kansu{\vp_\alpha}{k}
  +i\kakko{\kansu{p_\alpha}{M}-\kansu{p_\alpha}{1}}
  \kansu{\vp_\alpha}{M}
  \nonumber\\
  &&+
  i\kansu{p_\alpha}{1}\cdot2n_\alpha\pi
  -i\delt\mu_\alpha\wa{M}{k=1}\kansu{p_\alpha}{k}
  \bigg]
  \Bigg]\Bigg\vert_{\kansu{\vp_\alpha}{M}=\kansu{\vp_\alpha}{0}}
  \ .
\end{eqnarray}
By rewriting the $\vp$-integrals to the $\delta$-functions,
(\ref{gen:phyoji}) becomes
\begin{eqnarray}
  Z
  &=&
  \lim_{M\to\infty}e^{-iKc_{N+1}T}
  \prod^N_{\alpha=1}
  \Bigg[
  e^{{i\over2}\mu_\alpha T}
  \wa{\infty}{n_\alpha=-\infty}e^{-in_\alpha\pi}
  \int^{2\pi}_0{d\kansu{\vp_\alpha}{M}\over2\pi}
  \int^\infty_0\prod^M_{j=1}d\kansu{p_\alpha}{j}
  \nonumber\\
  &&\times
  \exp\bigg[
  i\kakko{\kansu{p_\alpha}{M}-\kansu{p_\alpha}{1}}
  \kansu{\vp_\alpha}{M}
  +i\kansu{p_\alpha}{1}\cdot2n_\alpha\pi
  -i\delt\mu_\alpha\wa{M}{k=1}\kansu{p_\alpha}{k}
  \bigg]
  \nonumber\\
  &&\times
  \prod^{M-1}_{l=1}
  \kansu{\delta}{\kansu{p_\alpha}{l+1}-\kansu{p_\alpha}{l}}
  \Bigg]
  \nonumber\\
  &=&
  \lim_{M\to\infty}
  e^{-iKc_{N+1}T}
  \prod^N_{\alpha=1}
  \bigg[
  e^{{i\over2}\mu_\alpha T}
  \wa{\infty}{n_\alpha=-\infty}
  e^{-in_\alpha\pi}\int^\infty_0d\kansu{p_\alpha}{M}
  e^{i\kansu{p_\alpha}{M}
    \kakko{2n_\alpha\pi-\mu_\alpha T}}
  \bigg]\ ,
\end{eqnarray}
where $\kansu{p_\alpha}{k}$- and $\kansu{\vp_\alpha}{M}$-integrals
have been made in the second equality.
Performing the $\kansu{p_\alpha}{M}$-integrals
(We regularize $i\kansu{p_\alpha}{M}(2n_\alpha\pi-\mu_\alpha T)
\to i\kansu{p_\alpha}{M}(2n_\alpha\pi-\mu_\alpha T+i\ve)$ for
the $\kansu{p_\alpha}{M}$-integrals to converge.), we obtain
\begin{equation}
  Z
  =
  e^{-iKc_{N+1}T}
  \prod^N_{\alpha=1}
  \bigg[
  e^{{i\over2}\mu_\alpha T}{1\over i}
  \wa{\infty}{n_\alpha=-\infty}
  {e^{i2n_\alpha\pi\cdot{1\over2}}
    \over2n_\alpha\pi+\mu_\alpha T}
  \bigg]\ ,
\end{equation}
where we have changed $n_\alpha\to-n_\alpha$.
By applying the formula (see Appendix~\ref{furoku:wakoushiki});
\begin{equation}\label{cpn:wakoushiki}
  \wa{\infty}{n=-\infty}
  {e^{i2n\pi\ve}\over2n\pi+\vp}
  =
  {ie^{-i\vp\ve}\over1-e^{-i\vp}}\ ,\
  \kakko{0<\ve<1}\ ,
\end{equation}
the final form becomes
\begin{equation}
  \label{ncp:genmitsu}
  Z
  =
  {e^{-iKc_{N+1}T}\over
      \prod^N_{\alpha=1}\kakko{1-e^{-i\mu_\alpha T}}}\ .
\end{equation}

\subsection{The WKB Approximation}
\label{sec:multi:wkb}

{}From the $\theta$-expression of the trace formula (\ref{ncp:ato}),
we read the solutions of the equations of motion as
\begin{equation}
  \kansu{\theta_\alpha}{k}=0\ ,
\end{equation}
for all $k$ with the $\vp_\alpha$'s are arbitrary.
We consider the WKB approximation as the large $K$ expansion.
We therefore make a change of variables such that
\begin{equation}
  \kansu{\theta_\alpha}{k}
  =
  0+\kansu{x_\alpha}{k}/\sqrt{K}\ .
\end{equation}
The leading order term of the expansion becomes
\begin{eqnarray}
  Z_0
  &\equiv&
  \lim_{M\to\infty}
  e^{-iKc_{N+1}T}
  \prod^N_{\alpha=1}
  \Bigg[
  e^{i{3\over2}\mu_\alpha T}
  \wa{\infty}{n_\alpha=-\infty}
  e^{-i3n_\alpha\pi}
  \int^{2\pi}_0
  {d\kansu{\vp_\alpha}{M}\over2\pi}
  \nonumber\\
  &&\times
  \int^\infty_{-\infty}\prod^M_{i=1}
  {d\kansu{\vp_\alpha}{i}\over2\pi}
  \int^\infty_0\prod^M_{j=1}
  {\kansu{x_\alpha}{j}d\kansu{x_\alpha}{j}\over K}
  \nonumber\\
  &&\times
  \exp\bigg[
  i\wa{M}{k=1}
  \kakko{1+{\kansu{x_\alpha}{k}^2\over2K}}
  \kakko{\kansu{\vp_\alpha}{k}
    \kansu{\vp_\alpha}{k-1}-\delt\mu_\alpha}
  \bigg]
  \Bigg]\ .
\end{eqnarray}
Putting
\begin{equation}
  \kansu{y_\alpha}{k}
  =
  {\kansu{x_\alpha}{k}^2\over2K}\ ,
\end{equation}
we obtain
\begin{eqnarray}\label{gen:tenkainoato}
  Z_0
  &=&
  \lim_{M\to\infty}
  e^{-iKc_{N+1}T}
  \prod^N_{\alpha=1}
  \Bigg[
  e^{{i\over2}\mu_\alpha T}
  \wa{\infty}{n_\alpha=-\infty}
  e^{-in_\alpha\pi}\int^{2\pi}_0
  {d\kansu{\vp_\alpha}{M}\over2\pi}
  \int^\infty_{-\infty}
  \prod^{M-1}_{i=1}
  {d\kansu{\vp_\alpha}{i}\over2\pi}
  \nonumber\\
  &&\times
  \int^\infty_0
  \prod^M_{j=1}
  d\kansu{y_\alpha}{j}
  \exp\bigg[
  i\wa{M}{k=1}\kansu{y_\alpha}{k}
  \kakko{\kansu{\vp_\alpha}{k}-\kansu{\vp_\alpha}{k-1}
  -\delt\mu_\alpha}
  \bigg]
  \Bigg]
  \nonumber\\
  &=&
  \lim_{M\to\infty}
  e^{-iKc_{N+1}T}
  \prod^N_{\alpha=1}
  \Bigg[
  \wa{\infty}{n_\alpha=-\infty}
  \int^{2\pi}_0{d\kansu{\vp_\alpha}{M}\over2\pi}
  \int^\infty_{-\infty}\prod^{M-1}_{i=1}
  {d\kansu{\vp_\alpha}{i}\over2\pi}
  \int^\infty_0\prod^M_{j=1}d\kansu{y_\alpha}{j}
  \nonumber\\
  &&\times
  \exp\bigg[i\wa{M}{k=1}
  \kakko{\kansu{y_\alpha}{k}-{1\over2}}
  \kakko{\kansu{\vp_\alpha}{k}-\kansu{\vp_\alpha}{k-1}
  -\delt\mu_\alpha}
  \bigg]
  \Bigg]\Bigg\vert_{\kansu{\vp_\alpha}{M}=\kansu{\vp_\alpha}{0}}
  \ .
\end{eqnarray}
Comparing (\ref{gen:tenkainoato}) and the $p$-expression of the
trace formula (\ref{gen:phyoji}), we find that they are quite the same.
Thus without any explicit calculations we obtain the result of the WKB
approximation;
\begin{equation}
  Z_0
  =
  {e^{-iKc_{N+1}T}\over
      \prod^N_{\alpha=1}\kakko{1-e^{-i\mu_\alpha T}}}\ ,
\end{equation}
and conclude that {\em the WKB approximation gives the
exact result}.

\section{Symplectic 2-Forms on $CP^N$ and $CQ^N$ Revisited}
\label{sec:symplectic}

So far we have dealt with quantum mechanics.
In this section we revisit the (classical) symplectic 2-forms
on $CP^N$ and $CQ^N$.
Especially we construct that on $CQ^N$ with the unitary form
to comprehend symplectic 2-forms in a unified manner with
that on $CP^N$.

$CP^N$ is defined by
\begin{eqnarray}
  \label{cpn:teigi}
  CP^N
  &\equiv&
  \left\{
    P\in\kansu{M}{N+1;C}\Big\vert
    P^2=P,P^\dagger=P,\tr P=1
  \right\}
  \nonumber\\
  &=&
  \left\{
    U\pmatrix{1&&&\cr&0&&\cr&&\ddots&\cr&&&0\cr}U^{-1}
    \Big\vert U\in\kansu{U}{N+1}
  \right\}
  \nonumber\\
  &\cong&
  {\kansu{U}{N+1}\over \kansu{U}{1}\times\kansu{U}{N}}\ .
\end{eqnarray}
$CP^N$ is connected and  has $N+1$ local chart:
\begin{equation}
  CP^N
  =
  U_1\cup U_2\cup\cdots\cup U_{N+1}\ .
\end{equation}
An element $P_1$ on $U_1$ is
\begin{eqnarray}
  \label{cpn:pnokatachi}
  P_1
  &\equiv&
  \tilde{\fxi}
  \kakko{\tilde{\fxi}^\dagger\tilde{\fxi}}^{-1}
  \tilde{\fxi}^\dagger
  \nonumber\\
  &=&
  {1\over1+\fxi^\dagger\fxi}
  \pmatrix{
    1&\fxi^\dagger\cr
    \fxi&\fxi\fxi^\dagger\cr
    }\ ,
  \nonumber\\
  &&
  \tilde{\fxi}
  \equiv\pmatrix{1\cr\fxi\cr}\ ,\
  \fxi\in{\bf C}^N\ ,
\end{eqnarray}
and $P_\alpha$ on $U_\alpha (\alpha=2,\cdots,N+1)$ is obtained from
$P_1$ as
\begin{eqnarray}
  P_\alpha
  &=&
  AP_1A^{-1}\ ,
  \nonumber\\
  A
  &\equiv&
  \pmatrix{
    0&&&&1&&&&\cr
    &1&&&0&&&&\cr
    &&\ddots&&\vdots&&&&\cr
    &&&1&&&&&\cr
    1&0&\cdots&&0&&&&\cr
    &&&&&1&&\cr
    &&&&&&&\ddots&\cr
    &&&&&&&&1\cr
    }
  \ .
\end{eqnarray}
The symplectic 2-form on $CP^N$ is defined by
\begin{equation}
  \label{twoform:cpnteigi}
  \omega_{CP^N}
  \equiv
  \tr PdP\wedge dP\ ,
\end{equation}
where it should be noted that this expression does not
depend on any $\alpha$.
$P$ in (\ref{cpn:pnokatachi}) is rewritten by
\begin{eqnarray}
  \label{cpn:pwake}
  P
  &=&
  \pmatrix{1&-\fxi^\dagger\cr\fxi&1_N}
  \pmatrix{1&\cr&0_N\cr}
  \pmatrix{1&-\fxi^\dagger\cr\fxi&1_N\cr}^{-1}\ ,
  \nonumber\\
  &&
  \pmatrix{1&-\fxi^\dagger\cr\fxi&1_N\cr}^{-1}
  =
  \pmatrix{\kakko{1+\fxi^\dagger\fxi}^{-1}&\cr
    &\kakko{1_N+\fxi\fxi^\dagger}^{-1}\cr}
  \pmatrix{1&\fxi^\dagger\cr-\fxi&1_N\cr}\ ,
\end{eqnarray}
and then $dP$ becomes
\begin{equation}
  \label{cpn:pwaked}
  dP
  =
  \pmatrix{1&-\fxi^\dagger\cr\fxi&1_N}
  \pmatrix{0&\kakko{1+\fxi^\dagger\fxi}^{-1}d\fxi^\dagger\cr
    \kakko{1_N+\fxi\fxi^\dagger}^{-1}d\fxi&0_N\cr}
  \pmatrix{1&-\fxi^\dagger\cr\fxi&1_N\cr}^{-1}\ .
\end{equation}
Thus, substituting (\ref{cpn:pwake}) and (\ref{cpn:pwaked}) into
(\ref{twoform:cpnteigi}), we obtain the explicit form of
the symplectic 2-form as
\begin{eqnarray}
  \label{twoform:cpn}
  \omega_{CP^N}
  &=&
  \kansu{\tr}{
    \pmatrix{1&-\fxi^\dagger\cr\fxi&1_N\cr}
    \pmatrix{\kakko{1+\fxi^\dagger\fxi}^{-1}d\fxi^\dagger
      \wedge\kakko{1_N+\fxi\fxi^\dagger}^{-1}d\fxi\cr
      &0\cr}
    \pmatrix{1&-\fxi^\dagger\cr\fxi&1_N\cr}^{-1}
    }
  \nonumber\\
  &=&
  \kakko{1+\fxi^\dagger\fxi}^{-1} d\fxi^\dagger
  \wedge
  \kakko{1_N+\fxi\fxi^\dagger}^{-1}d\fxi
  \nonumber\\
  &=&
  {1\over1+\fxi^\dagger\fxi}
  d\fxi^\dagger
  \wedge
  \kakko{1_N-{\fxi\fxi^\dagger\over1+\fxi^\dagger\fxi}}
  d\fxi\ ,
\end{eqnarray}
where we have used the relation
\begin{equation}
  \kakko{1_N+\fxi\fxi^\dagger}^{-1}
  =
  1_N-\fxi\kakko{1+\fxi^\dagger\fxi}^{-1}\fxi^\dagger
  =
  1_N-{\fxi\fxi^\dagger\over1+\fxi^\dagger\fxi}\ ,
\end{equation}
in the last equality.
The volume form on $CP^N$ is then
\begin{equation}
  \Omega_{CP^N}
  =
  {\omega_{CP^N}^N\over N!}
  =
  {\prod^N_{\alpha=1}d\xi_\alpha^*d\xi_\alpha
    \over
    \kakko{1+\fxi^\dagger\fxi}^{N+1}}\ .
\end{equation}

$CQ^N$ is defined by
\begin{equation}
  \label{cqn:teigi}
  CQ^N
  \equiv
  \left\{
    Q\in\kansu{M}{N+1;C}\vert
    Q^2=Q,
    \eta Q^\dagger\eta=Q,
    \tr Q=1
  \right\}\ .
\end{equation}
$CQ^N$ consists of two connected components;
\begin{eqnarray}
  CQ^N
  &=&
  CQ^N_+\cup CQ^N_-\ ,
  \nonumber\\
  CQ^N_+
  &\equiv&
    \left\{
    V\pmatrix{1&&&\cr&0&&\cr&&\ddots&\cr&&&0\cr}V^{-1}
    \Bigg\vert
    V\in\kansu{U}{1,N}
  \right\}
  \nonumber\\
  &\cong&
  {\kansu{U}{1,N}\over\kansu{U}{1}\times\kansu{U}{N}}
  \cong
  \kansu{D^N}{C}
  \equiv
  \left\{
    \ffz\in{\bf C}^N\vert\ffz^\dagger\ffz<1
  \right\}\ ,
  \nonumber\\
  CQ^N_-
  &\equiv&
    \left\{
    V\pmatrix{0&&&\cr&1&&\cr&&\ddots&\cr&&&0\cr}V^{-1}
    \Bigg\vert
    V\in\kansu{U}{1,N}
  \right\}\ ,
\end{eqnarray}
where
\begin{equation}
  \label{cqn:uteigi}
  \kansu{U}{1,N}
  \equiv
  \left\{
    V\in\kansu{M}{N+1;C}\vert V^\dagger\eta V=\eta
  \right\}\ ,
\end{equation}
and
\begin{equation}
  \label{cqn:etateigi}
  \eta
  \equiv
  \kansu{\rm diag}{1,-1,\cdots,-1}\ .
\end{equation}

We should note that although we have used, in section
\ref{sec:multi:noncompact},
$\eta=\kansu{\rm diag}{1,\cdots,1,-1}$ for consistency
with the previous work~\cite{WESHS},
we use (\ref{cqn:etateigi}) in this section.
Hereafter we consider $CQ^N_+$ part only and we write
$CQ^N_+$ as $CQ^N$.
$CQ^N$ has only one local chart
\begin{equation}
  Q
  =
  {1\over1-\fxi^\dagger\fxi}
  \pmatrix{1&-\fxi^\dagger\cr\fxi&-\fxi\fxi^\dagger\cr}\ ,
\end{equation}
where
\begin{equation}
  \fxi\in D^N\ .
\end{equation}
The symplectic 2-form on $CQ^N$ is defined by
\begin{equation}
  \label{twoform:cqnteigi}
  \omega_{CQ^N}
  \equiv
  \kansu{\tr}{QdQ\wedge dQ}\ .
\end{equation}
By a similar way with the $CP^N$ case,
we calculate (\ref{twoform:cqnteigi}).
$Q$ is rewritten by
\begin{equation}
  Q
  =
  \pmatrix{1&\fxi^\dagger\cr\fxi&1_N\cr}
  \pmatrix{1&\cr&0_N\cr}
  \pmatrix{1&\fxi^\dagger\cr\fxi&1_N\cr}^{-1}\ ,
\end{equation}
and
\begin{equation}
  dQ
  =
  \pmatrix{1&\fxi^\dagger\cr\fxi&1_N\cr}
  \pmatrix{0&\kakko{1-\fxi^\dagger\fxi}^{-1}d\fxi^\dagger\cr
    \kakko{1_N-\fxi\fxi^\dagger}^{-1}d\fxi&0_N\cr}
  \pmatrix{1&\fxi^\dagger\cr\fxi&1_N\cr}^{-1}\ .
\end{equation}
Thus the symplectic 2-form becomes
\begin{eqnarray}
  \label{twoform:cqn}
  \omega_{CQ^N}
  &=&
  \kansu{\tr}{
    \pmatrix{1&\fxi^\dagger\cr\fxi&1_N\cr}
    \pmatrix{\kakko{1-\fxi^\dagger\fxi}^{-1}d\fxi^\dagger
      \wedge\kakko{1_N-\fxi\fxi^\dagger}^{-1}d\fxi&\cr
      &0_N\cr}
    \pmatrix{1&\fxi^\dagger\cr\fxi&1_N\cr}^{-1}
    }
  \nonumber\\
  &=&
  \kakko{1-\fxi^\dagger\fxi}^{-1}d\fxi^\dagger
  \wedge
  \kakko{1_N-\fxi\fxi^\dagger}^{-1}d\fxi
  \nonumber\\
  &=&
  {1\over1-\fxi^\dagger\fxi}
  d\fxi^\dagger
  \wedge
  \kakko{1_N+{\fxi\fxi^\dagger\over1-\fxi^\dagger\fxi}}
  d\fxi\ ,
\end{eqnarray}
where we have used the relation
\begin{equation}
  \kakko{1_N-\fxi\fxi^\dagger}^{-1}
  =
  1_N+\fxi\kakko{1-\fxi^\dagger\fxi}^{-1}\fxi^\dagger
  =
  1_N+{\fxi\fxi^\dagger\over1-\fxi^\dagger\fxi}\ .
\end{equation}
Thus the volume form is
\begin{equation}
  \Omega_{CQ^N}
  =
  {\omega_{CQ^N}^N\over N!}
  =
  {\prod^N_{\alpha=1}d\xi_\alpha^*d\xi_\alpha
    \over
    \kakko{1-\fxi^\dagger\fxi}^{N+1}}\ .
\end{equation}

We have obtained the symplectic 2-forms on $CP^N$ and $CQ^N$.
The difference between them is sign of denominators.
The expression of $CP^N$ is unitary, while that of $CQ^N$
is not unitary.
Now we construct $\omega_{CQ^N}$ from unitary form with taking
account of the above difference.
We embed $CQ^N$ to $CP(l^2(C))$.
$CP(l^2(C))$ is defined by
\begin{equation}
  \label{cpl:teigi}
  \kansu{CP}{\kansu{l^2}{C}}
  \equiv
  \left\{
    P\in\kansu{M}{\kansu{l^2}{C}}
    \vert
    P^2=P,P^\dagger=P,\tr P=1
  \right\}\ ,
\end{equation}
where $l^2(C)$ is defined by
\begin{equation}
  \kansu{l^2}{C}
  \equiv
  \left\{
    \ffz\in C^\infty\vert\ffz^\dagger\ffz<\infty
  \right\}\ ,
\end{equation}
and $M(l^2(C))$ denotes whole matrices of bounded linear
operators from $l^2(C)$ to $l^2(C)$.
We put
\begin{eqnarray}
  \label{cqn:phat}
  \hat P
  \equiv
  \kansu{\hat P}{\hat{\fxi}}
  &\equiv&
  \pmatrix{1\cr\hat{\fxi}}
  \kakko{1+\hat{\fxi}^\dagger\hat{\fxi}}^{-1}
  \kakko{1\quad\hat{\fxi}^\dagger}
  \nonumber\\
  &=&
  {1\over1+\hat{\fxi}^\dagger\hat{\fxi}}
  \pmatrix{1&\hat{\fxi}^\dagger\cr
    \hat{\fxi}&\hat{\fxi}\hat{\fxi}^\dagger\cr}
  \nonumber\\
  &=&
  \pmatrix{1&-\hat{\fxi}^\dagger\cr
    \hat{\fxi}&1_\infty\cr}
  \pmatrix{1&\cr&0_\infty\cr}
  \pmatrix{1&-\hat{\fxi}\cr
    \hat{\fxi}&1_\infty\cr}^{-1}\ .
\end{eqnarray}
The symplectic 2-form is defined by
\begin{eqnarray}
  \label{twoform:mugenteigi}
  \hat\omega
  &\equiv&
  \kansu{\tr}{\hat Pd\hat P\wedge d\hat P}
  \nonumber\\
  &=&
  {1\over1+\hat{\fxi}^\dagger\hat{\fxi}}
  \kakko{
    d\hat{\fxi}^\dagger\wedge d\hat{\fxi}
    -
    {d\hat{\fxi}^\dagger\hat{\fxi}
      \wedge
      \hat{\fxi}^\dagger d\hat{\fxi}
      \over
      1+\hat{\fxi}^\dagger\hat{\fxi}
    }
  }\ ,
\end{eqnarray}
where the result of the $CP^N$ case, (\ref{twoform:cpn}),
 has been put in the second equality.
Hereafter we identify $l^2(C)$ as the Fock space generated from $C^N$;
\begin{equation}
  \kansu{l^2}{C}
  \cong
  C\oplus C^N\oplus C^N\otimes C^N
  \oplus\cdots\oplus\kakko{C^N}^{\otimes n}
  \oplus\cdots\ ,
\end{equation}
and assign
\begin{eqnarray}
  \hat{\fxi}
  \equiv
  \kakko{\fxi,\fxi\otimes\fxi,\cdots,\fxi^{\otimes n},\cdots}^T\ ,
  \nonumber\\
  \fxi\in D^N\ .
\end{eqnarray}
Then noting that
\begin{eqnarray}
  \hat{\fxi}^\dagger\hat{\fxi}
  &=&
  \fxi^\dagger\fxi
  +\kakko{\fxi^\dagger\fxi}^2
  +\cdots
  +\kakko{\fxi^\dagger\fxi}^n
  +\cdots
  \nonumber\\
  &=&
  {\fxi^\dagger\fxi\over1-\fxi^\dagger\fxi}\ ,
\end{eqnarray}
and
\begin{eqnarray}
  \hat{\fxi}^\dagger d\hat{\fxi}
  &=&
  {1\over\kakko{1-\fxi^\dagger\fxi}^2}\fxi^\dagger d\fxi\ ,
  \nonumber\\
  d\hat{\fxi}^\dagger\hat{\fxi}
  &=&
  {1\over\kakko{1-\fxi^\dagger\fxi}^2}d\fxi^\dagger\fxi\ ,
  \nonumber\\
  d\hat{\fxi}^\dagger\wedge d\hat{\fxi}
  &=&
  {1\over\kakko{1-\fxi^\dagger\fxi}^2}
  d\fxi^\dagger\wedge d\fxi
  +{2\over\kakko{1-\fxi^\dagger\fxi}^3}
  d\fxi^\dagger\fxi\wedge\fxi^\dagger d\fxi\ ,
\end{eqnarray}
we obtain the explicit form of (\ref{twoform:mugenteigi}) as
\begin{eqnarray}
  \label{cqn:univ:form}
  \hat\omega
  &=&
  {1\over1-\fxi^\dagger\fxi}
  \kakko{d\fxi^\dagger\wedge d\fxi
    +
    {d\fxi^\dagger\fxi\wedge\fxi^\dagger d\fxi
      \over
      1-\fxi^\dagger\fxi}
  }
  \nonumber\\
  &=&
  {1\over1-\fxi^\dagger\fxi}
  d\fxi^\dagger\wedge
  \kakko{1_N+{\fxi\fxi^\dagger\over1-\fxi^\dagger\fxi}}
  d\fxi\ .
\end{eqnarray}
This result is quite the same with (\ref{twoform:cqn}).
The symplectic 2-form on $CQ^N$ has been constructed
with the unitary expression.
We note that the similar discussion for $CP^N$ is also possible.
However the embedding of $CP^N$ to $CP(l^2(C))$ is trivial:
\begin{equation}
  \fxi\to\hat{\fxi}=\pmatrix{\fxi\cr0\cr\vdots\cr}\ .
\end{equation}

\section{Classical Partition Functions on $CP^N$ and $CQ^N$ Revisited}
  \label{sec:koten}

In this section we revisit the classical partition functions
of $CP^N$ and $CQ^N$.
We make the exact calculations of them directly.
We also calculate them by lifting to the Gaussian forms.

\subsection{The Partition Function of $CP^N$}

In this subsection we construct the partition function of $CP^N$
and calculate it exactly.
$CP^N$ is defined in (\ref{cpn:teigi}).

We construct the partition function of $CP^N$.
By putting
\begin{equation}
  \ffz
  =
  U\pmatrix{1\cr0\cr\vdots\cr0\cr}
  \equiv
  U{\bf e}_1\ ,
\end{equation}
an element of $CP^N$ becomes
\begin{equation}
  \label{cpn:uz}
  P
  =
  U\pmatrix{1&&&\cr&0&&\cr&&\ddots&\cr&&&0\cr}U^{-1}
  =
  \ffz\ffz^\dagger\ .
\end{equation}
Now we define
\begin{eqnarray}
  S_C^N
  &\equiv&
  \left\{
    \ffz\in C^{N+1}\vert
    \ffz^\dagger\ffz=1
  \right\}
  \nonumber\\
  &=&
  \left\{
    U{\bf e}_1\vert U\in\kansu{U}{N+1}
  \right\}
  \nonumber\\
  &\cong&
  {\kansu{U}{N+1}\over\kansu{U}{N}}\ ,
\end{eqnarray}
and a map
\begin{equation}
  \pi\ :\
  S_C^N\to CP^N\ ,\
  \kansu{\pi}{\ffz}\equiv\ffz\ffz^\dagger\ .
\end{equation}
(\ref{cpn:uz}) indicates that $\pi$ is an onto-mapping.
Also
\begin{equation}
  \kansu{\pi}{e^{i\theta}\ffz}
  =
  \kansu{\pi}{\ffz}\ ,
\end{equation}
holds.
Thus $\pi$ induces the principal $U(1)$ bundle.
\begin{equation}
  \kansu{U}{1}\to S_C^N\to CP^N\ .
\end{equation}
We define a Hamiltonian as
\begin{eqnarray}
    H
  &\equiv&
  \tr\kakko{Ph}
  \nonumber\\
  &=&
  \kansu{\tr}{\ffz\ffz^\dagger h}
  \nonumber\\
  &=&
  \ffz^\dagger h\ffz
  \qquad{\rm under}\ \ffz^\dagger\ffz=1
  \nonumber\\
  &=&
  \wa{N}{\alpha=0}\theta_\alpha\vert z_\alpha\vert^2\ ,
\end{eqnarray}
where
\begin{equation}
  \label{cpn:hamiltonian}
  h
  =
  \pmatrix{\theta_0&&&\cr&\theta_1&&\cr&&\ddots&\cr&&&\theta_N\cr}\ ,\
  \kakko{0<\theta_0<\theta_1<\cdots<\theta_N}\ .
\end{equation}
Thus the ``Gaussian form'' of the partition function of $CP^N$
is defined by
\begin{eqnarray}
  \label{cpn:gauss}
  Z
  &\equiv&
  \int{\prod^N_{\alpha=0}dz_\alpha^*dz_\alpha
    \over\pi^{N+1}}
  \int^\infty_{-\infty}{d\lambda\over2\pi}
  e^{-\rho H-i\lambda\kakko{\ffz^\dagger\ffz-1}}
  \nonumber\\
  &=&
  \int{\prod^N_{\alpha=0}dz_\alpha^*dz_\alpha\over\pi^{N+1}}
  \int^\infty_{-\infty}{d\lambda\over2\pi}
  \exp
  \bigg[
  -\rho\wa{N}{\alpha=0}
  \theta_\alpha\vert z_\alpha\vert^2
  -i\lambda\kakko{\wa{N}{\alpha=0}\vert z_\alpha\vert^2-1}
  \bigg]\ .
\end{eqnarray}

To examine the equivalence between (\ref{cpn:gauss}) and the standard
form of the partition function which is given by
\begin{equation}
  \int\Omega_{CP^N}
  e^{-\rho\kansu{\tr}{Ph}}\ ,
\end{equation}
we make a change of variables such that
\begin{eqnarray}
  \xi_\alpha
  &=&
  z_\alpha/z_0\ ,\
  \kakko{\alpha=1,\cdots,N}\ ,
  \nonumber\\
  \eta
  &=&
  z_0\kakko{\wa{N}{\alpha=0}\vert z_\alpha/z_0\vert^2}^{1/2}\ .
\end{eqnarray}
Equation (\ref{cpn:gauss}) then becomes
\begin{eqnarray}
  Z
  &=&
  \int^\infty_{-\infty}
  {d\lambda\over2\pi}
  \int_{{\bf C}^N}
  {\prod^N_{\alpha=1}d\xi_\alpha^*d\xi_\alpha
    \over
     \pi^N\kakko{1+\fxi^\dagger\fxi}^{N+1}}
   \int_{\bf C}
   {d\eta\over\pi}\left\vert\eta\right\vert^{2N}
   \nonumber\\
   &&\times
   \exp
   \bigg[
   -\rho
  {\theta_0+\wa{N}{\alpha=1}\theta_\alpha
    \left\vert\xi_\alpha\right\vert^2
    \over
    \kakko{1+\fxi^\dagger\fxi}^{N+1}}
  -i\lambda\kakko{\left\vert\eta\right\vert^2-1}
  \bigg]\ .
\end{eqnarray}
After integrating the angular part of $\eta$
and writing the $\lambda$-integral as the $\delta$-function,
we obtain
\begin{eqnarray}
  \label{cpn:ato}
  Z
  &=&
  \int_{{\bf C}^N}
  {\prod^N_{\alpha=1}d\xi_\alpha^*d\xi_\alpha
    \over
    \pi^N\kakko{1+\fxi^\dagger\fxi}^{N+1}}
  \int^\infty_0
  dvv^N\kansu{\delta}{v-1}
  \exp
  \bigg[
  -\rho v
  {\theta_0+\wa{N}{\alpha=1}
    \theta_\alpha\left\vert\xi_\alpha\right\vert^2
    \over
    1+\fxi^\dagger\fxi}
  \bigg]
  \nonumber\\
  &=&
  \int_{{\bf C}^N}
  {\prod^N_{\alpha=1}d\xi_\alpha^*d\xi_\alpha
    \over
    \pi^N\kakko{1+\fxi^\dagger\fxi}^{N+1}}
  \exp
  \bigg[
  -\rho
  {\theta_0+\wa{N}{\alpha=1}
    \theta_\alpha\left\vert\xi_\alpha\right\vert^2
    \over
    1+\fxi^\dagger\fxi}
  \bigg]\ ,
\end{eqnarray}
where we have put $\eta=\sqrt{v}e^{i\phi}$.
This is of course the standard form of the classical partition
function of $CP^N$.

By the DH theorem, we evaluate~\cite{WESHS} this integral,(\ref{cpn:ato}), to
\begin{equation}
  \label{cpn:dh}
  Z
  =
  \wa{N}{\alpha=0}
  {e^{-\rho\theta_\alpha}
    \over
    \rho^N\prod^N_{\beta=0\atop\beta\ne\alpha}
    \kakko{\theta_\beta-\theta_\alpha}}\ .
\end{equation}

We calculate (\ref{cpn:ato}) exactly.
Making a change of variables such that
\begin{equation}
  \xi_\alpha
  =
  \sqrt{u_\alpha}e^{i\vp_\alpha}\ ,
\end{equation}
and performing $\vp_\alpha$-integrals, we obtain
\begin{eqnarray}
  Z
  &=&
  \int^\infty_0
  {\prod^N_{\alpha=1}du_\alpha
    \over
    \kakko{1+\wa{N}{\alpha=1}u_\alpha}^{N+1}}
  \exp
  \bigg[
  -\rho
  {\theta_0+\wa{N}{\alpha=1}\theta_\alpha u_\alpha
    \over
    1+\wa{N}{\alpha=1}u_\alpha}
  \bigg]\ ,
  \nonumber\\
  &=&
  e^{-\rho\theta_0}
  \int^\infty_0
  {\prod^N_{\alpha=1}du_\alpha
    \over
    \kakko{1+\wa{N}{\alpha=1}u_\alpha}^{N+1}}
  \exp
  \bigg[
  -\rho
  {\wa{N}{\alpha=1}\kakko{\theta_\alpha-\theta_0}u_\alpha
    \over
    1+\wa{N}{\alpha=1}u_\alpha}
  \bigg]\ .
\end{eqnarray}
We will obtain (\ref{cpn:dh}) by mathematical induction for $N$.
Changing variables such that
\begin{eqnarray}
  u_\alpha
  &=&
  t_\alpha u_N\ ,\ \kakko{\alpha=1,\cdots,N-1}\ ,
  \nonumber\\
  x
  &=&
  \wa{N}{\alpha=1}u_\alpha\ ,
  \nonumber\\
  &&
  {\kansu{\partial}{u_1,\cdots,u_N}
    \over
    \kansu{\partial}{t_1,\cdots,t_{N-1},x}}
  =
  {x^{N-1}
    \over
    \kakko{1+\wa{N-1}{\alpha=1}t_\alpha}^N}\ ,
\end{eqnarray}
leads to
\begin{eqnarray}
  &&\hspace{-10mm}
  \kansu{Z_{N+1}}{\rho;\theta_0,\theta_1,\cdots,\theta_N}
  \nonumber\\
  &=&
  e^{-\rho\theta_0}
  \int^\infty_0
  {x^{N-1}dx\over\kakko{1+x}^{N+1}}
  \int^\infty_0
  {\prod^N_{\alpha=1}dt_\alpha
    \over
    \kakko{1+\wa{N-1}{\alpha=1}t_\alpha}^N}
  \nonumber\\
  &&\times
  \exp
  \bigg[
  -\rho{x\over1+x}
  {\kakko{\theta_N-\theta_0}
    +\wa{N-1}{\alpha=1}
    \kakko{\theta_\alpha-\theta_0}t_\alpha
    \over
    1+\wa{N-1}{\alpha=1}t_\alpha}
  \bigg]
  \nonumber\\
  &=&
  e^{-\rho\theta_0}
  \int^\infty_0
  {x^{N-1}dx\over\kakko{1+x}^{N+1}}
  \kansu{Z_N}
  {\rho{x\over1+x};\theta_1-\theta_0,
    \cdots,\theta_N-\theta_0}\ ,
\end{eqnarray}
where we have written $Z$ as
$\kansu{Z_{N+1}}{\rho;\theta_0,\theta_1,\cdots,\theta_N}$
to emphasize dependency on the parameters.
By assumption we put
\begin{eqnarray}
  &&\hspace{-10mm}
  \kansu{Z_N}
  {\rho{x\over1+x};\theta_1-\theta_0,
    \cdots,\theta_N-\theta_0}
  \nonumber\\
  &=&
  \wa{N}{\alpha=1}
  {e^{-\rho{x\over1+x}\kakko{\theta_\alpha-\theta_0}}
      \over
      \kakko{\rho{x\over1+x}}^{N-1}
      \prod^N_{\beta=1\atop\beta\ne\alpha}
      \left\{
        \kakko{\theta_\beta-\theta_0}
        -\kakko{\theta_\alpha-\theta_0}
      \right\}}\ ,
\end{eqnarray}
to obtain
\begin{eqnarray}
  &&\hspace{-10mm}
  \kansu{Z_{N+1}}{\rho;\theta_0,\theta_1,\cdots,\theta_N}
  \nonumber\\
  &=&
  {e^{-\rho\theta_0}\over\rho^{N-1}}
  \wa{N}{\alpha=1}
  {1\over\prod^N_{\beta=1\atop\beta\ne\alpha}
    \kakko{\theta_\beta-\theta_\alpha}}
  \int^\infty_0
  {dx\over\kakko{1+x}^2}
  e^{-\rho{x\over1+x}\kakko{\theta_\alpha-\theta_0}}
  \nonumber\\
  &=&
  {e^{-\rho\theta_0}\over\rho^{N-1}}
  \wa{N}{\alpha=1}
  {1\over\prod^N_{\beta=1\atop\beta\ne\alpha}
    \kakko{\theta_\beta-\theta_\alpha}}
  {1-e^{-\rho\kakko{\theta_\alpha-\theta_0}}
    \over\rho\kakko{\theta_\alpha-\theta_0}}
  \nonumber\\
  &=&
  {1\over\rho^N}
  \Bigg[
  -\wa{N}{\alpha=1}
  {e^{-\rho\theta_0}
    \over
    \prod^N_{\beta=0\atop\beta\ne\alpha}
    \kakko{\theta_\beta-\theta_\alpha}}
  +
  \wa{N}{\alpha=1}
  {e^{^\rho\theta_0}
    \over
    \prod^N_{\beta=0\atop\beta\ne\alpha}
    \kakko{\theta_\beta-\theta_\alpha}}
  \Bigg]\ .
\end{eqnarray}
Noting the relation (see Appendix \ref{sec:shoumei}),
\begin{equation}
  \label{cpn:shikihenkei}
  \wa{N}{\alpha=1}
  {1
    \over
    \prod^N_{\beta=0\atop\beta\ne\alpha}
    \kakko{\theta_\beta-\theta_\alpha}}
  =
  -{1
    \over
    \prod^N_{\beta=1}
    \kakko{\theta_\beta-\theta_0}}\ ,
\end{equation}
we finally obtain
\begin{eqnarray}
  \label{cpn:genmitsu}
  &&\hspace{-10mm}
  \kansu{Z_{N+1}}{\rho;\theta_1,\cdots,\theta_0}
  \nonumber\\
  &=&
  {1\over\rho^N}
  \Bigg[
  {e^{-\rho\theta_0}
    \over
    \prod^N_{\beta=1}
    \kakko{\theta_\beta-\theta_0}}
  +
  \wa{N}{\alpha=1}
  {e^{-\rho\theta_0}
    \over
    \prod^N_{\beta=0\atop\beta\ne\alpha}
    \kakko{\theta_\beta-\theta_\alpha}}
  \Bigg]
  \nonumber\\
  &=&
  {1\over\rho^N}
  \wa{N}{\alpha=0}
  {e^{-\rho\theta_\alpha}
    \over
    \prod^N_{\beta=0\atop\beta\ne\alpha}
    \kakko{\theta_\beta-\theta_\alpha}}\ .
\end{eqnarray}
Equation (\ref{cpn:genmitsu}) coincides with the result of the
DH theorem.
In other words the DH theorem holds in the partition function
of $CP^N$.

Next we calculate (\ref{cpn:gauss}).
To obtain (\ref{cpn:ato}), we perform the $\lambda$-integral
in (\ref{cpn:gauss}).
Now we perform the $\ffz$-integrals, which are the Gaussian,
before the $\lambda$-integrals to obtain
\begin{eqnarray}
  \label{cpn:gaussato}
  Z
  &=&
  \int^\infty_{-\infty}
  {d\lambda\over2\pi}
  e^{-i\lambda}
  \int{\prod^N_{\alpha=0}dz_\alpha^*dz_\alpha\over\pi^{N+1}}
  \exp
  \bigg[
  -\wa{N}{\alpha=0}
  \kakko{\rho\theta_\alpha+i\lambda}
  \vert z_\alpha\vert^2
  \bigg]
  \nonumber\\
  &=&
  \int^\infty_{-\infty}
  {d\lambda\over2\pi}
  e^{i\lambda}
  \prod^N_{\alpha=0}
  {1\over\rho\theta_\alpha+i\lambda}\ .
\end{eqnarray}
(\ref{cpn:gaussato}) is easily calculated by contour
integral to be
\begin{eqnarray}
  \label{cpn:ryuusuu}
  Z
  &=&
  \wa{N}{\alpha=0}
  {\rm Res}_{z=i\rho\theta_\alpha}
  e^{iz}
  \prod^N_{\beta=0}
  {1\over\rho\theta_\beta+iz}
  \nonumber\\
  &=&
  \wa{N}{\alpha=0}
  e^{-\rho\theta_\alpha}
  \prod^N_{\beta=0\atop\beta\ne\alpha}
  {1\over\rho\kakko{\theta_\beta-\theta_\alpha}}\ .
\end{eqnarray}
Of course this result coincides with (\ref{cpn:genmitsu}).
We note that (\ref{cpn:genmitsu}) (and (\ref{cpn:ryuusuu}))
can be written by the determinant form
\begin{equation}
  \label{cpn:detform}
  Z
  =
  {1\over\rho^N}
  \left\vert
    \matrix{
      e^{-\rho\theta_0}& e^{-\rho\theta_1}&\cdots& e^{-\rho\theta_N}\cr
      1&1&\cdots&1\cr
      \theta_0&\theta_1&\cdots&\theta_N\cr
      \vdots&\vdots&\ddots&\vdots\cr
      \theta_0^{N-2}&\theta_1^{N-2}&\cdots&\theta_N^{N-2}\cr
    }
  \right\vert
  \Bigg/
  \left\vert
    \matrix{
      1&1&\cdots&1\cr
      \theta_0&\theta_1&\cdots&\theta_N\cr
      \theta_0^2&\theta_1^2&\cdots&\theta_N^2\cr
      \vdots&\vdots&\ddots&\vdots\cr
      \theta_0^{N-1}&\theta_1^{N-1}&\cdots&\theta_N^{N-1}\cr
    }
  \right\vert\ .
\end{equation}

\subsection{The Partition Function of $CQ^N$}

In this subsection we construct the partition function of $CQ^N$ and
calculate it exactly.

$CQ^N$ is defined in (\ref{cqn:teigi}).
By putting
\begin{equation}
  \ffz
  =
  V\pmatrix{1\cr0\cr\vdots\cr0\cr}
  =
  V{\bf e}_1\ ,
\end{equation}
an element of $CQ^N$ becomes
\begin{equation}
  \label{cqn:yousokakikae}
  V\pmatrix{1&&&\cr&0&&\cr&&\ddots&\cr&&&0\cr}V^{-1}
  =
  V\pmatrix{1&&&\cr&0&&\cr&&\ddots&\cr&&&0\cr}V^\dagger\eta
  =
  \ffz\ffz^\dagger\eta\ ,
\end{equation}
where we have used $V^{-1}=\eta V^\dagger\eta$ in (\ref{cqn:uteigi}).
Then we define
\begin{eqnarray}
  Q_C^N
  &\equiv&
  \left\{
    \ffz\in{\bf C}^{N+1}\vert\ffz^\dagger\eta\ffz=1
  \right\}
  \nonumber\\
  &=&
  \left\{
    V{\bf e}_1\vert V\in\kansu{U}{1,N}
  \right\}
  \nonumber\\
  &\cong&
  {\kansu{U}{1,N}\over\kansu{U}{N}}\ ,
\end{eqnarray}
and a map
\begin{equation}
  \pi\ :\
  Q_C^N\to CQ^N,\
  \kansu{\pi}{\ffz}\equiv\ffz\ffz^\dagger\eta\ .
\end{equation}
{}From (\ref{cqn:yousokakikae}), we find that $\pi$ is onto-mapping.
Also
\begin{equation}
  \kansu{\pi}{e^{i\theta}\ffz}
  =
  \kansu{\pi}{\ffz}\ ,
\end{equation}
holds.
Thus $\pi$ induces the principal $U(1)$ bundle;
\begin{equation}
  \kansu{U}{1}\to
  Q_C^N\to
  CQ^N\ .
\end{equation}

We define a Hamiltonian as
\begin{eqnarray}
  \label{cqn:hamiltonian}
  H
  &\equiv&
  \tr Qh
  \nonumber\\
  &=&
  \kansu{\tr}{\ffz\ffz^\dagger\eta h}
  \nonumber\\
  &=&
  \ffz^\dagger\eta h\ffz
  \qquad{\rm under}\ \ffz^\dagger\eta\ffz=1\ ,
\end{eqnarray}
where we have put
\begin{equation}
  Q
  \equiv
  \ffz\ffz^\dagger\eta\ ,
\end{equation}
and
\begin{equation}
  \label{cqn:hamgyouretsu}
  h
  =
  \pmatrix{\theta_0&&&\cr&\theta_1&&&\cr&&\ddots&\cr&&& \theta_N\cr}\ ,\
  \kakko{\theta_0>\theta_1>\cdots>\theta_N>0}\ .
\end{equation}
We should pay attention to the ordering of $\theta_\alpha$'s.
(See (\ref{cpn:hamiltonian}).)
The explicit form of (\ref{cqn:hamiltonian}) is
\begin{equation}
  H
  =
  \theta_0\vert z_0\vert^2
  -\wa{N}{\alpha=1}\theta_\alpha\vert z_\alpha\vert^2\ .
\end{equation}
Thus the ``Gaussian form'' of the partition function of $CQ^N$
is defined by
\begin{eqnarray}
  \label{cqn:ato}
  Z
  &\equiv&
  \int{\prod^N_{\alpha=0}dz_\alpha^*dz_\alpha\over\pi^{N+1}}
  \int^\infty_{-\infty}{d\lambda\over2\pi}
  e^{-\rho H+i\lambda\kakko{\ffz^\dagger\eta\ffz-1}}
  \nonumber\\
  &=&
  \int{\prod^N_{\alpha=0}dz_\alpha^*dz_\alpha\over\pi^{N+1}}
  \int^\infty_{-\infty}{d\lambda\over2\pi}
  e^{-\rho\kakko{\theta_0\vert z_0\vert^2
      -\wa{N}{\alpha=1}\theta_\alpha\vert z_\alpha\vert^2}
    +i\lambda\kakko{
      \vert z_0\vert^2
      -\wa{N}{\alpha=1}\vert z_\alpha\vert^2-1}
    }\ .
\end{eqnarray}

To examine the equivalence between (\ref{cqn:ato}) and the standard form,
we make a change of variables such that
\begin{eqnarray}
  \zeta
  &=&
  z_0
  \sqrt{1-\wa{N}{\alpha=1}\vert z_\alpha/z_0\vert^2}\ ,
  \nonumber\\
  \xi_\alpha
  &=&
  z_\alpha/z_0\ ,\
  \kakko{\alpha=1,\cdots,N}\ .
\end{eqnarray}
Equation (\ref{cqn:ato}) then becomes
\begin{eqnarray}
  Z
  &=&
  \int_{D^N}
  {\prod^N_{\alpha=1}d\xi_\alpha^*d\xi_\alpha
    \over
    \pi^N\kakko{1-\fxi^\dagger\fxi}^{N+1}}
  \int{d\zeta^*d\zeta\over\pi}
  \kansu{\delta}{\vert\zeta\vert^2-1}
  \nonumber\\
  &&\times
  \exp
  \bigg[
  -\rho\vert\zeta\vert^2
  {\theta_0-\wa{N}{\alpha=1}\theta_\alpha\xi_\alpha^*\xi_\alpha
    \over
    1-\fxi^\dagger\fxi}
  \bigg]\ .
\end{eqnarray}
Putting $\zeta=\sqrt{r}e^{i\vp}$ and integrating
with respect to $r$ and $\vp$, we obtain
\begin{equation}
  \label{cqn:kotenato}
  Z
  \equiv
  \int_{D^N}
  {\prod^N_{\alpha=1}d\xi_\alpha^*d\xi_\alpha
    \over
    \pi^N\kakko{1-\fxi^\dagger\fxi}^{N+1}}
  \exp
  \bigg[
  -\rho
  {\theta_0-\wa{N}{\alpha=1}
    \theta_\alpha\xi_\alpha^*\xi_\alpha
    \over1-\fxi^\dagger\fxi}
  \bigg]\ .
\end{equation}
This is the standard form of the classical partition function of $CQ^N$.

Now we calculate (\ref{cqn:kotenato}) directly.
Making a change of variables such that
\begin{equation}
  \xi_\alpha
  =
  \sqrt{u_\alpha}e^{i\vp_\alpha}\ ,
\end{equation}
and performing $\vp_\alpha$-integrals, we obtain
\begin{eqnarray}
  Z
  &=&
  \int_{\wa{N}{\alpha=1}u_\alpha<1}
  {\prod^N_{\alpha=1}du_\alpha
    \over
    \kakko{1-\wa{N}{\alpha=1}u_\alpha}^{N+1}}
  \exp
  \bigg[
  -\rho
  {\theta_0-\wa{N}{\alpha=1}\theta_\alpha u_\alpha
    \over
    1-\wa{N}{\alpha=1}u_\alpha}
  \bigg]
  \nonumber\\
  &=&
  e^{-\rho\theta_0}
  \int_{\wa{N}{\alpha=1}u_\alpha<1}
  {\prod^N_{\alpha=1}du_\alpha
    \over
    \kakko{1-\wa{N}{\alpha=1}u_\alpha}^{N+1}}
  \exp
  \bigg[
  -\rho
  {\wa{N}{\alpha=1}
    \kakko{\theta_0-\theta_\alpha}u_\alpha
    \over
    1-\wa{N}{\alpha=1}u_\alpha}
  \bigg]\ .
\end{eqnarray}
Then putting
\begin{eqnarray}
  x_\alpha
  =
  {u_\alpha\over1-\wa{N}{\alpha=1}u_\alpha}\ ,\
  \kakko{1\le\alpha\le N}\ ,
  \nonumber\\
  {\kansu{\partial}{u_1,\cdots,u_N}
    \over
    \kansu{\partial}{x_1,\cdots,x_N}}
  =
  {1\over\kakko{1+\wa{N}{\alpha=1}x_\alpha}^{N+1}}\ ,
\end{eqnarray}
leads to
\begin{eqnarray}
  \label{cqn:kekka}
  Z
  &=&
  e^{-\rho\theta_0}
  \int^\infty_0
  \prod^N_{\alpha=1}dx_\alpha
  \exp
  \bigg[
  -\rho\wa{N}{\alpha=1}
  \kakko{\theta_0-\theta_\alpha}x_\alpha
  \bigg]
  \nonumber\\
  &=&
  {e^{-\rho\theta_0}
    \over
    \prod^N_{\alpha=1}
    \rho\kakko{\theta_0-\theta_\alpha}}\ .
\end{eqnarray}
This is the exact result, which corresponds to (\ref{cpn:genmitsu})
in $CP^N$ case.

Next we perform the $z_0$-integral as the Gaussian in (\ref{cqn:ato});
\begin{eqnarray}
  Z
  &=&
  \int^\infty_{-\infty}
  {d\lambda\over2\pi}
  {e^{-i\lambda}\over\rho\theta_0-i\lambda}
  \prod^N_{\alpha=1}\int
  {dz_\alpha^*dz_\alpha\over\pi}
  e^{z_\alpha^*\kakko{\rho\theta_\alpha-i\lambda}z_\alpha}
  \nonumber\\
  &=&
  e^{-\rho\theta_0}
  \prod^N_{\alpha=1}
  \int^\infty_{-\infty}
  {dz_\alpha^*dz_\alpha\over\pi}
  e^{-\rho z_\alpha^*\kakko{\theta_0-\theta_\alpha}z_\alpha}\ ,
\end{eqnarray}
where we have made residue calculations with respect to
$\lambda$ in the second equality.
Noting that $\theta_0-\theta_\alpha>0$, we perform the
$z_\alpha$-integral to find
\begin{equation}
  \label{cqn:zkara}
  Z
  =
  e^{-\rho\theta_0}
  {1\over\prod^N_{\alpha=1}\rho\kakko{\theta_0-\theta_\alpha}}\ .
\end{equation}
This result corresponds to (\ref{cpn:ryuusuu}) in $CP^N$ case
and, of course, coincides with (\ref{cqn:kekka}).
We note that, as similar with $CP^N$ case, (\ref{cpn:detform}),
(\ref{cqn:kekka}) (or (\ref{cqn:zkara})) can be written
by the determinant form
\begin{equation}
  Z
  =
  {\kakko{-1}^N\over\rho^N}
  \left\vert
    \matrix{
      e^{-\rho\theta_0}&0&\cdots&0\cr
      1&1&\cdots&1\cr
      \theta_0&\theta_1&\cdots&\theta_N\cr
      \vdots&\vdots&\ddots&\vdots\cr
      \theta_0^{N-2}&\theta_1^{N-2}&\cdots&\theta_N^{N-2}\cr
    }
  \right\vert
  \Bigg/
  \left\vert
    \matrix{
      1&1&\cdots&1\cr
      \theta_0&\theta_1&\cdots&\theta_N\cr
      \theta_0^2&\theta_1^2&\cdots&\theta_N^2\cr
      \vdots&\vdots&\ddots&\vdots\cr
      \theta_0^{N-1}&\theta_1^{N-1}&\cdots&\theta_N^{N-1}\cr
    }
  \right\vert\ .
\end{equation}

\subsection{The Partition Function of $CQ^N$ from the ``Universal''
 Partition Function of $CP(l^2(C))$}
\label{sec:univ}

In this subsection we construct the partition function of $CQ^N$
by means of the notion developed in the last part of the
previous section, that is, the embedding of $CQ^N$ to $CP(l^2(C))$.
$CP(l^2(C))$ is defined in (\ref{cpl:teigi}) and the symplectic
2-form is in (\ref{twoform:mugenteigi}).
Since the ``Liouville measure'', $\lim_{N\to\infty}{\hat\omega}^N/N!$,
diverges, there needs some regularization.
If it is possible, the ``universal'' partition function
is defined by
\begin{equation}
  \label{univ:ptteigi}
  \int_{\kansu{CP}{\kansu{l^2}{C}}}
  \kansu{dv_{\kansu{CP}{\kansu{l^2}{C}}}}{\hat P}
  e^{-\rho\kansu{\tr}{\hat P\hat H}}\ ,
\end{equation}
where $dv_{CP(l^2(C))}$ is a regularized measure.
Unfortunately we do not know whether (\ref{univ:ptteigi})
can be defined, however we can define the pullback to $CQ^N$.
The definition of $CQ^N$ is given in (\ref{cqn:teigi})
and the symplectic 2-form of $CQ^N$ is given in (\ref{twoform:cqnteigi}).
Thus we obtain
\begin{equation}
  \label{univ:taiseki}
  {\omega_{CQ^N}^N\over N!}
  =
  {{\hat\omega_{\kansu{CP}{\kansu{l^2}{C}}}}^N\over N!}\ ,
\end{equation}
by (\ref{cqn:univ:form}).
Hamiltonian of $CQ^N$ is given in (\ref{cqn:hamgyouretsu}).
We define a Hamiltonian of $CP(l^2(C))$ from that of $CQ^N$ such that
\begin{equation}
  \hat H
  =
  \pmatrix{
    \theta_0&&&&\cr
    &2\theta_01_N-\tilde \theta&&&\cr
    &&\ddots&&\cr
    &&&\kakko{n\theta_01_N-\kakko{n-1}\tilde \theta}\otimes
    \overbrace{1_N\otimes\cdots\otimes1_N}^{n-2}&\cr
    &&&&\ddots\cr
  }\ ,
\end{equation}
where
\begin{equation}
  \tilde \theta
  =
  \pmatrix{
    \theta_1&&\cr
    &\ddots&\cr
    &&\theta_N\cr
  }\ .
\end{equation}
Using the explicit form of $\hat P$ in (\ref{cqn:phat})
\begin{eqnarray}
  \hat P
  &=&
  {1\over1+\hat{\fxi}^\dagger\hat{\fxi}}
  \pmatrix{
    1&\hat{\fxi}^\dagger\cr
    \hat{\fxi}&\hat{\fxi}\hat{\fxi}^\dagger
  }
  \nonumber\\
  &=&
  \kakko{1-\fxi^\dagger\fxi}
  \pmatrix{
    1&&&&&\cr
    &\fxi\fxi^\dagger&&&*&\cr
    &&\fxi\fxi^\dagger\otimes\fxi\fxi^\dagger&&&\cr
    &&&\ddots&&\cr
    &**&&&\kakko{\fxi\fxi^\dagger}^{\otimes\kakko{n-1}}&\cr
    &&&&&\ddots\cr
  }\ ,
  \nonumber\\
  &&
  \pmatrix{
    \kakko{1+\hat{\fxi}^\dagger\hat{\fxi}}^{-1}
    =
    1-\fxi^\dagger\fxi\ ,\cr
    *,**\ {\rm denote\ unnecessary\ elements\ for\ later\ calculation}\cr
  }\ ,
\end{eqnarray}
we obtain
\begin{eqnarray}
  \kansu{\tr}{\hat P\hat H}
  &=&
  \kakko{1-\fxi^\dagger\fxi}
  \left[
    \theta_0+\wa{\infty}{n=2}
    \left\{
      n\theta_0\fxi^\dagger\fxi-\kakko{n-1}\fxi^\dagger\tilde \theta\fxi
    \right\}
    \kakko{\fxi^\dagger\fxi}^{n-2}
  \right]
  \nonumber\\
  &=&
  \kakko{1-\fxi^\dagger\fxi}
  \kakko{\theta_0-\fxi^\dagger\tilde \theta\fxi}
  \wa{\infty}{n=1}n\kakko{\fxi^\dagger\fxi}^{n-1}
  \nonumber\\
  &=&
  {\theta_0-\fxi^\dagger\tilde \theta\fxi\over1-\fxi^\dagger\fxi}
  \nonumber\\
  &=&
  {\theta_0-\wa{N}{\alpha=1}\theta_\alpha\vert\xi_\alpha\vert^2
    \over
    1-\wa{N}{\alpha=1}\vert\xi_\alpha\vert^2}\ .
\end{eqnarray}
This is nothing but $\tr(QH)$ on $CQ^N$.
Thus we obtain the partition function;
\begin{equation}
  Z
  \equiv
  \int_{D^N}
  {\prod^N_{\alpha=1}d\xi_\alpha^*d\xi_\alpha
    \over
    \pi^N\kakko{1-\fxi^\dagger\fxi}^{N+1}}
  \exp
  \bigg[
  -\rho
  {\theta_0-\wa{N}{\alpha=1}
    \theta_\alpha\xi_\alpha^*\xi_\alpha
    \over1-\fxi^\dagger\fxi}
  \bigg]\ ,
\end{equation}
which is just the partition function of $CQ^N$, (\ref{cqn:kotenato}).

\section{Discussion}
\label{sec:giron}

We have examined the WKB-exactness of the trace formula of
some representation of $U(N,1)$ in terms of the multi-periodic
coherent state as an extension of our previous works.
We will be able to examine the WKB-exactness of trace formulas
in some representations of more complex manifolds such as the flag
manifold in some coherent states.
However all of the coherent states will not lead to the
WKB-exactness although it holds in the same Hamiltonian with some coherent
states.
It is interesting that what classes of coherent states lead
to the WKB-exactness.

We have performed the exact calculations of the classical partition
functions of $CP^N$ and $CQ^N$ directly and by lifting to the Gaussian forms.
These discussions will be extended to the Grassmann manifold
straightforwardly~\cite{Fujii}.
Also it may be applied to the flag manifold.

\begin{center}
  {\bf Acknowledgments}
\end{center}

K. Fujii wishes to thank A. Asada for useful discussions.

K. Fujii was partially supported by the Grant-in-Aid for Scientific
Research on Priority Areas, No.08211103.

\begin{center}
  {\Large\bf Appendix}
\end{center}

\appendix

\section{The Proof of (\protect\ref{cpn:wakoushiki})}\label{furoku:wakoushiki}

We prove the formula,
\begin{equation}
  \wa{\infty}{n=-\infty}
  {e^{i2n\pi\ve}\over2n\pi+\vp}
  =
  {ie^{-i\vp\ve}\over1-e^{-i\vp}}\ ,\
  \kakko{0<\ve<1}\ .
\end{equation}
Consider the function
\begin{equation}
  \kansu{f}{x}
  =
  e^{-i\vp x}\ ,\
  \kakko{0<x<1}\ ,
\end{equation}
and expand it as the Fourier series:
\begin{equation}
  \label{ftenkai}
  e^{-i\vp x}
  =
  \wa{\infty}{n=-\infty}
  f_ne^{i2\pi nx}\ .
\end{equation}
Its coefficient is then read as
\begin{eqnarray}
  f_n
  &=&
  \int^1_0dxe^{-i2\pi nx}e^{-i\vp x}
  \nonumber\\
  &=&
  {1-e^{-i\vp}\over i\kakko{2\pi n+\vp}}\ .
\end{eqnarray}
Thus (\ref{ftenkai}) is
\begin{equation}
  e^{-i\vp x}
  =
  {1-e^{-i\vp}\over i}
  \wa{\infty}{n=-\infty}{e^{i2\pi nx}\over2\pi n+\vp}\ ,
\end{equation}
and then
\begin{equation}
  \label{koshiki}
  \wa{\infty}{n=-\infty}
  {e^{i2\pi nx}\over2\pi n+\vp}
  =
  {ie^{-i\vp x}\over1-e^{-i\vp}}\ .
\end{equation}
If we put $x=\ve$ in (\ref{koshiki}), we just obtain the formula.
Also we can write (\ref{koshiki}) to the form
\begin{equation}
  \wa{\infty}{n=-\infty}
  {e^{i2\pi nx}\over2\pi n+\vp}
  =
  {e^{i\kakko{1/2-x}\vp}\over2\sin{\vp\over2}}\ ,
\end{equation}
which is the formula used in \cite{FKNS}.

\section{The Proof of (\protect\ref{cpn:shikihenkei})}\label{sec:shoumei}

We define symbols;
\begin{eqnarray}
  \Delta
  &\equiv&
  \prod^N_{\alpha=0}
  \prod^{\alpha-1}_{\beta=0}
  \kakko{\theta_\alpha-\theta_\beta}\ ,
  \nonumber\\
  \kansu{\Delta}{\alpha}
  &\equiv&
  \prod^N_{\beta=0\atop\beta\ne\alpha}
  \prod^{\beta-1}_{\gamma=0\atop\gamma\ne\alpha}
  \kakko{\theta_\beta-\theta_\gamma}\ ,
\end{eqnarray}
where $\Delta$ is the $N+1$-th Vandermonde's determinant
and $\kansu{\Delta}{\alpha}$ is $N$-th one omitted $\theta_\alpha$.
Then the relation
\begin{equation}
  \prod^N_{\beta=0\atop\beta\ne\alpha}
  \kakko{\theta_\beta-\theta_\alpha}
  =
  \kakko{-1}^\alpha
  {\Delta\over\kansu{\Delta}{\alpha}}\ ,
\end{equation}
holds.
Thus the left-hand side of (\ref{cpn:shikihenkei}) is
\begin{eqnarray}
  \wa{N}{\alpha=1}
  {1
    \over
    \prod^N_{\beta=0\atop\beta\ne\alpha}
    \kakko{\theta_\beta-\theta_\alpha}}
  &=&
  \wa{N}{\alpha=0}
  {1
    \over
    \prod^N_{\beta=0\atop\beta\ne\alpha}
    \kakko{\theta_\beta-\theta_\alpha}}
  -{1
    \over
    \prod^N_{\beta=1}
    \kakko{\theta_\beta-\theta_0}}
  \nonumber\\
  &=&
  \wa{N}{\alpha=0}
  {\kakko{-1}^\alpha\kansu{\Delta}{\alpha}\over\Delta}
  -{1
    \over
    \prod^N_{\beta=1}
    \kakko{\theta_\beta-\theta_0}}
  \nonumber\\
  &=&
  {1\over\Delta}
  \left\vert
    \matrix{
      1              & 1              & \cdots & 1              \cr
      1              & 1              & \cdots & 1              \cr
      \theta_0       & \theta_1       & \cdots & \theta_N       \cr
      \theta_0^2     & \theta_1^2     & \cdots & \theta_N^2     \cr
      \vdots         & \vdots         & \ddots & \vdots         \cr
      \theta_0^{N-1} & \theta_1^{N-1} & \cdots & \theta_N^{N-1} \cr}
  \right\vert
  -{1
    \over
    \prod^N_{\beta=1}\kakko{\theta_\beta-\theta_0}}
  \nonumber\\
  &=&
  -{1
    \over
    \prod^N_{\beta=1}\kakko{\theta_\beta-\theta_0}}
  \nonumber\\
  &=&
  \kakko{\rm r.h.s.\ of\ (\ref{cpn:shikihenkei})}\ .
\end{eqnarray}

\section{The Proof of the Closedness of $\omega$}
\label{sec:close}

{}From the property of projection, $P^2=P$,
the symplectic 2-form is written by
\begin{equation}
  \omega
  \equiv
  \kansu{\tr}{PdP\wedge dP}
  =
  \kansu{\tr}{PdP\wedge dPP}\ .
\end{equation}
The exterior derivative of it then becomes
\begin{eqnarray}
  \label{close:gaibibun}
  d\omega
  &=&
  2\kansu{\tr}{PdP\wedge dP\wedge dP}
  \nonumber\\
  &=&
  2\kansu{\tr}{PdP\wedge dP\wedge dPP}\ ,
\end{eqnarray}
where $P^2=P$ has been used in the second equality.
Further multiplying $P$ to $dP=PdP+dPP$, which is the exterior
derivative of $P=P^2$, we obtain
\begin{equation}
  \label{close:kankeiichi}
  PdPP=0\ ,
\end{equation}
and its derivative,
\begin{equation}
  \label{close:kankeini}
  PdP\wedge dP
  =
  dP\wedge dPP\ .
\end{equation}
Then (\ref{close:gaibibun}) becomes
\begin{eqnarray}
  d\omega
  &=&
  2\kansu{\tr}{dP\wedge dP\wedge PdPP}
  \nonumber\\
  &=&
  0\ ,
\end{eqnarray}
where (\ref{close:kankeini}) has been used in the first equality
and (\ref{close:kankeiichi}) has been applied in the second equality.
Thus we conclude that the symplectic 2-form $\omega$ is close.
Note that the proof is global since no any local charts do not appeared
in it.


\begin{thebibliography}{99}
\bibitem{Stone}
  M. Stone,
  Nucl. Phys. {\bf B314} (1989) 557.
\bibitem{Rajeev}
  S. G. Rajeev, S. K. Rama, and S. Sen,
  J. Math. Phys. {\bf 35} (1994) 2259.
\bibitem{PCS}
  R. F. Picken,
  J. Phys. A. (Math. Gen) {\bf 22} (1989) 2285.\\
  L. S. Schulman,
  Phys. Rev. {\bf 176} (1968) 1558.\\
  J. S. Dowker,
  J. Phys. {\bf A3} (1970) 451.
\bibitem{Blau}
  M. Blau,
  Int. J. Mod. Phys. {\bf A} (1991) 365.
\bibitem{DH}
  J.J. Duistermaat and G. J. Heckman,
  Invent. Math. {\bf 69} (1982) 259,
  {\it ibid.} {\bf 72} (1983) 153.
\bibitem{Atiyah}
  M.F. Atiyah, Asterisque {\bf 131} (1985) 43.
\bibitem{FKSF1}
  K. Funahashi, T. Kashiwa, S. Sakoda and K. Fujii,
  J. Math. Phys. {\bf 36} (1995) 3232.
\bibitem{WESHS}
  K. Funahashi, T. Kashiwa, S. Sakoda and K. Fujii,
  J. Math. Phys. {\bf 36} (1995) 4590.
\bibitem{FKS}
  K. Fujii, T. Kashiwa and S. Sakoda,
  J. Math. Phys. {\bf 37} (1996) 567.
\bibitem{Perelomov}
  A. Perelomov,
  {\em Generalized Coherent States and Their Applications}
  (Springer-Verlag, Berlin, 1986).
\bibitem{Schwinger}
  J. Schwinger,
  {\em ON ANGULAR MOMENTUM} in {\em QUANTUM THEORY OF ANGULAR MOMENTUM}
  (Academic press, New York, 1965).
\bibitem{NR}
  H. B. Nielsen and D. Rohrlich,
  Nucl. Phys. {\bf B299} (1988) 471.
\bibitem{TK}
  T. Kashiwa,
  Int. J. Mod. Phys. {\bf A5} (1990) 375.
\bibitem{FKNS}
  K. Funahashi, T. Kashiwa, S. Nima and S. Sakoda,
  Nucl. Phys. {\bf B453} (1995) 508.
\bibitem{KFKF}
  K. Fujii and K. Funahashi,
  to be published in J. Math. Phys.
\bibitem{RFP}
  R. F. Picken,
  J. Math. Phys. {\bf 31} (1990) 616.
\bibitem{Fujii}
  K. Fujii,
  in preparation.
\end{thebibliography}
\end{document}